\documentclass[doublecol, a4paper, showpacs]{epl2}

\usepackage{graphicx}
\usepackage{amsmath, amssymb}
\usepackage{lmodern}
\usepackage[T1]{fontenc}
\usepackage[utf8]{inputenc}
\usepackage[english]{babel}
\usepackage{float}
\usepackage{color}

\title{Hamming distance and mobility behavior in generalized rock-paper-scissors models}

\author{D. Bazeia\inst{1}\thanks{{Corresponding author; \email{bazeia@fisica.ufpb.br}}} \and J. Menezes{\inst{2}\,\! \inst{3}} \and B.F. de Oliveira \inst{4}\and J.G.G.S. Ramos\inst{1}}
\shortauthor{D.Bazeia \etal}

\institute{
	\inst{1}Departamento de F\'\i sica, Universidade Federal da Para\'\i ba, 58051-970 Jo\~ao Pessoa, Para\'\i ba, Brazil\\ 
	\inst{2}{Escola de Ci\^encias e Tecnologia, Universidade Federal do Rio Grande do Norte\\ Caixa Postal 1524, 59072-970, Natal, Rio Grande do Norte, Brazil}\\
	\inst{3}Institute for Biodiversity and Ecosystem Dynamics, University of Amsterdam, Science Park 904, 1098 XH Amsterdam, The Netherlands\\
	\inst{4}Departamento de F\'isica, Universidade Estadual de Maring\'a, 87020-900 Maring\'a, Paran\'a, Brazil
}

\abstract{This work reports on two related investigations of stochastic simulations which are widely used to study biodiversity and other related issues. We first deal with the behavior of the Hamming distance under the increase of the number of species and the size of the lattice, and then investigate how the mobility of the species contributes to jeopardize biodiversity. The investigations are based on the standard rules of reproduction, mobility and predation or competition, which are described by specific rules, guided by generalization of the rock-paper-scissors game, valid in the case of three species. The results on the Hamming distance indicate that it engenders universal behavior, independently of the number of species and the size of the square lattice. The results on the mobility confirm the prediction that it may destroy diversity, if it is increased to higher and higher values.}

\begin{document}

\maketitle

%%%%%%%%%%%%%%%%%%%%%%%%%%%%%%%%%%%%%%%%%%%%%%%%%%%%%%%%%%%%%%%%%%%%%%
\section{Introduction}

Despite the many ways to study complex systems, stochastic simulations represent an important tool that has been largely used to investigate collective behavior in nature. The procedure is based on a set of simple rules that are assessed randomly and, in particular, has been employed to model and understand biodiversity in nature; see, e.g., Refs.~\cite{b1,b2,ML,rps,kd,DL,Sch,n02,n04,n07,S} and references therein.

In this work we deal with generalized rock-paper-scissors models to describe stochastic network simulations of the May and Leonard type \cite{ML,DL}. We consider a square lattice of size $N\times N$ and follow three recent investigations, the first \cite{n02} describing how local dispersal may promote biodiversity in a real-life game, in which the predictions based on the rock-paper-scissors rules are empirically tested using a non-transitive model community containing three populations of {\it Escherichia coli}, the second \cite{n07} suggesting that population mobility may be central feature to describe real ecosystems, and the third \cite{SR} that shows how to use the Hamming distance \cite{H} to unveil the presence of chaos in complex biological systems. These studies develop simulations that engender cyclic competition, modeled as in the rock-paper-scissors game, which is controlled by the simple rules where paper wraps rock, rock crushes scissors and scissors cut paper \cite{rps}, and the generalizations that are illustrated in Fig.~\ref{fig1}.

An interesting behavior of the stochastic simulations is the formation of spiral patterns, which indicate a subjacent law that was recently investigated in \cite{SR}. The study unveiled the presence of chaos as an important component to guide evolution of natural life, taking the Hamming distance \cite{H} as a way to measure the difference between the time evolution corresponding to two slightly distinct initial states, via an algorithm developed in \cite{SR}. In the current work we further explore the Hamming distance concept, to confirm that it acquires the universal qualitative behavior identified in Ref.~\cite{SR}. To do this, we consider three distinct lattice sizes and several distinct systems, labeled by $n$, the first with three distinct species $a$, $b$, and $c$, with $n=3$, the second with four distinct species, $a$, $b$, $c$, and $d$, with $n=4$, the third with five distinct species, $a$, $b$, $c$, $d$, and $e$, with $n=5$, and so on until the case with $n=10$, respectively. We identify the species with the colors red ($a$), blue $(b)$, yellow $(c)$, green $(d)$ and magenta $(e)$, etc, and the empty sites with the color white.

The investigation starts describing how to implement the stochastic simulations, and then calculating the Hamming distance density for three lattice sizes and several distinct species. We then go on and study how mobility can be used to jeopardize biodiversity \cite{n07}, and how the Hamming distance can be related to the disappearance of biodiversity in the current context. We close the work with some comments and conclusions.

%%%%%%%%%%%%%%%%%%%%%%%%%%%%%%%%%%%%%%%%%%%%%%%%%%%%%%%%%%%%%%%%%%%%%%
\section{Stochastic simulations}
\label{sec2}

One supposes that the system evolves according to the three basic rules of mobility $(m)$, reproduction $(r)$, and predation or competition $(p)$, with $m+p+r=1$. The evolution is implemented according to the standard stochastic simulations, considering a square lattice in which the species and the empty sites (identified as $v$, with the color white) are equally but randomly distributed in the lattice, such that the quantity of individuals of each species (including the empty sites) is $L_n=N^2/(n+1)$ at the initial state, for $n=3, 4$ or $5$, representing the system with three, four or five species, respectively. We deal with three distinct lattices, with $N=250, 500$, and $1000$, implementing the numerical simulations with periodic boundary conditions. A site is considered active if it is occupied by an individual of one of the active species, and it may interact with one of its eight nearest neighbors, the Moore neighborhood.

\begin{figure}[t]
\centering
\includegraphics[scale=0.8]{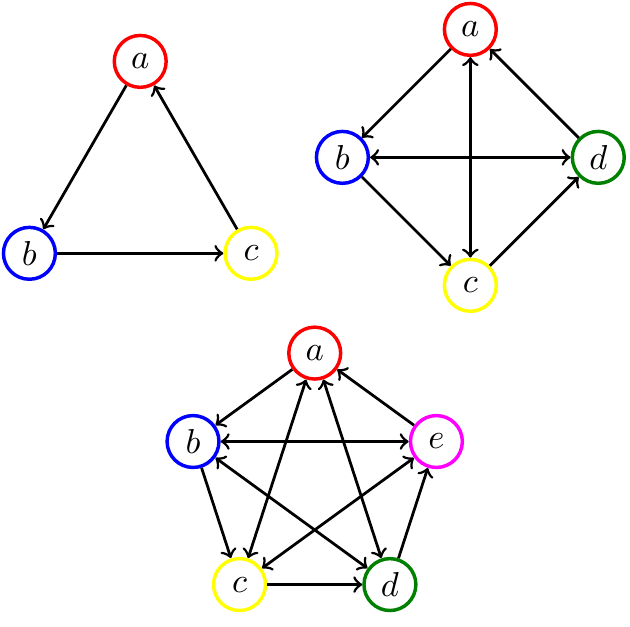}
\caption{The rock-paper-scissors rules for the system with three distinct species, and its generalizations to the cases of four and five species. The unidirectional and bidirectional arrows display the predation or competition rules.}
\label{fig1}
\end{figure}

We work with distinct systems, and implement the stochastic simulations considering similar environments, in which if $i$ stands for an active species and $v$ for an empty site, then reproduction is described by $i\,v\to i\,i$, and mobility by $i\,j\to j\,i$ or $i\,v\to v\,i$. The other rule, predation or competition, follows the rock-paper-scissors rules; however, since one enlarges the number of species, one needs to generalize the rules, and here we consider the cases explained in Fig.~\ref{fig1}. There one sees that the external arrows are all unidirectional, and the internal ones are bidirectional; that is, the first neighbors predate unidirectionally, and the second ones compete bidirectionally. To exemplify the unidirectional and the bidirectional behavior of the rule for predation or competition, we consider the system with five species, for instance. In this case, under the $p$ rule one gets: $a\, b \,\to\, a \,v$ and $b\, a\,\to\, b\, a$, but $a\, c\,\to\, a\, v$ and $c\, a\,\to\, c\, v$. Moreover, we study the Hamming distance density considering systems with three, four, five, six, seven, eight, nine and ten distinct species, with the generalization to a larger number of species following as suggested in Fig.~\ref{fig1}: the first neighbors interact unidirectionally, but all the others, the second, third, fourth, and fifth neighbors, when they exist, interact bidirectionally. We will also take $m=0.5$, $r=0.25$, and $p=0.25$ as the typical set of values for $(m,r,p)$, but we will vary $m$ increasing it to higher and higher values, to investigate how mobility contributes to end diversity, as first pointed out in Ref.~\cite{n07}. Anyway, in this work we will always consider $p=r$, for simplicity.

% fig 1 %%%%%%%%%%%%%%%%%%%%%
% fig 1 %%%%%%%%%%%%%%%%%%%%%

The dynamical process starts with the initial state, which we describe below, and with a random access to the square lattice, followed by a random selection of one of the three rules and by a random choice of one of the eight neighbors. It is then checked if the chosen site is active or empty: if it is empty, one returns to the lattice to simulate another access to it; if it contains an individual of one of the species, one takes the selected rule and uses it with the selected neighbor. We will describe the time evolution using generation, which is the time spent to access the lattice $N^2$ times. The stochastic simulations follow the standard procedure, and we implement the random accesses to the square lattice using the MT19937 generator of Makoto Matsumoto and Takuji Nishimura, which generates random real numbers in the interval $[0,1)$ with uniform distributioni \cite{gsl}.

% fig 2 %%%%%%%%%%%%%%%%%%%%%
\begin{figure}[t]
\centering
\includegraphics[scale=0.84]{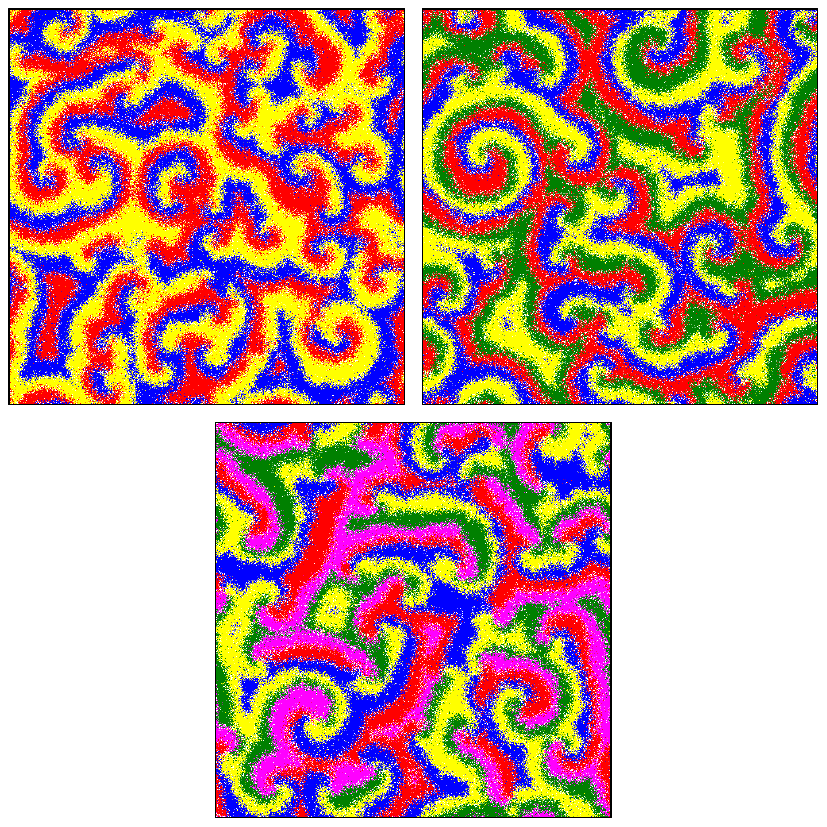}
\caption{Snapshots of the time evolution of the three systems, with three, four and five species, depicted in the top left, top right and bottom panels, respectively. The snapshots were taken after 10000 generations.}
\label{fig2}
\end{figure}
% fig 2 %%%%%%%%%%%%%%%%%%%%%

\begin{figure}[t]
\centering
\includegraphics{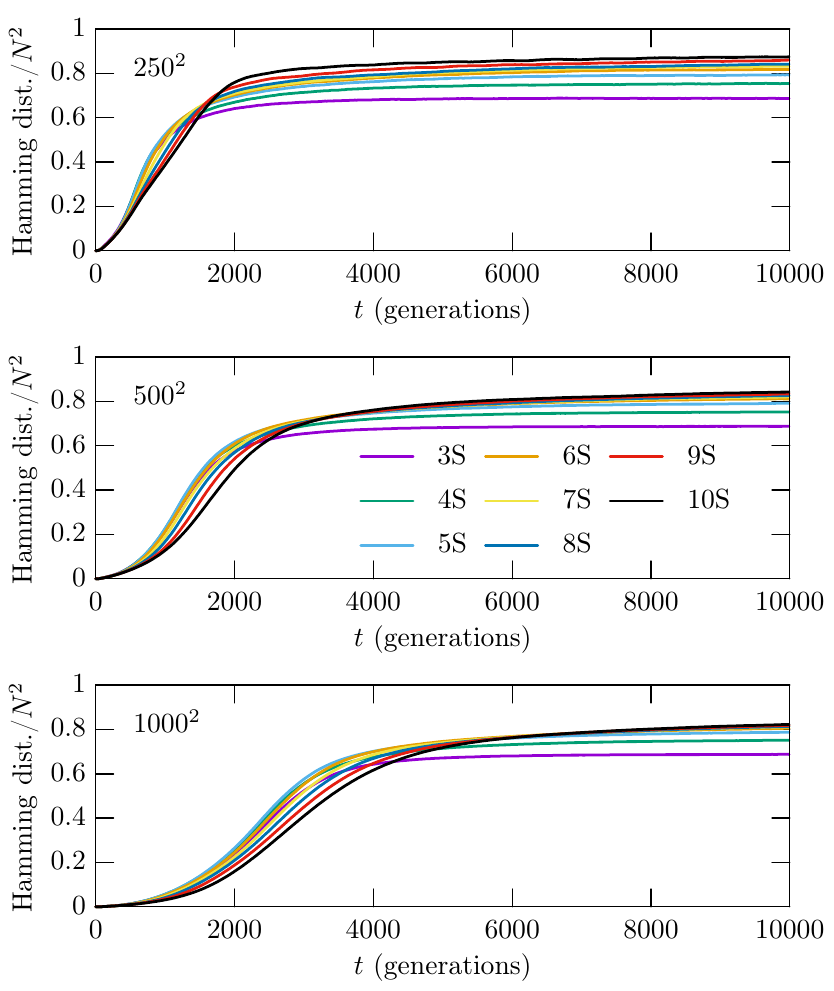}
\caption{The Hamming distance density is displayed as the average of a set of 1000 simulations. One shows the systems with three distinct lattice sizes, with $N=250, 500,$ and $1000$, and considers the cases with three, four, five, six, seven, eight, nine and ten species.}
\label{fig3}
\end{figure}

To prepare the initial state, one randomly chooses one among the $n$ species and the empty site with the same probability, and distributes it in the square lattice, repeating the procedure $N^2$ times, evenly filling all the sites of the square lattice. This gives a typical initial state, in which one has $N^2/(n+1)$ sites colored with the color white to represent the empty sites, and the same quantity for each one of the colors used to identify the $n$ species in the system. One uses this initial state to run the stochastic simulations to get the final configuration which is displayed in Fig.~\ref{fig2}, one panel for each one of the systems with three, four and five species that are displayed in the figure. There one sees that the system evolves forming specific patterns, in which the species form spirals and organize themselves in spatial portions of the lattice. This is known in the literature, and has been explored in a diversity of contexts to study biodiversity; see, e.g., Ref.~\cite{AB} and references therein. In particular, in all the snapshots that appear in Fig.~\ref{fig2}, a small fraction of the sites are empty sites, painted with the color white. Since the empty sites are passive sites, they are much less numerous than the colored species and do not aggregate, so they are almost invisible in the snapshots shown in Fig.~\ref{fig2}. We could also prepare the initial state with no empty sites, just distributing the $n$ species in the lattice, each one in this case with $N^2/n$ sites colored with the respective colors. This adds no qualitative modification in the results displayed in Figs.~\ref{fig2} and \ref{fig3}.

To further explore the stochastic network simulations, one investigates how the abundance or the density of individuals $l_i=L_i/N^2$ evolves in time. The results show that the size density of each one of the species fluctuates around the same average value, which depends on the number of species in the system. They do not add to unit because of the presence of empty sites, that fluctuates around a lower value, since they are passive sites. The study is similar to the case of three distinct species investigated before in \cite{SR}, so we omit the details in the current work.

% fig 3 %%%%%%%%%%%%%%%%%%%%%
% fig 3 %%%%%%%%%%%%%%%%%%%%%

%%%%%%%%%%%%%%%%%%%%%%%%%%%%%%%%%%%%%%%%%%%%%%%%%%%%%%%%%%%%%%%%%%%%%%
\section{Hamming Distance}
\label{sec3}

Let us now investigate the Hamming distance \cite{H} between the two lattices that describe the two evolutions that start with slightly distinct initial states. We follow the procedure introduced in \cite{SR}, recalling that it was used as a tool to unveil the presence of chaos in the stochastic simulations that one usually implements to investigate biodiversity in nature. Here we return to it to show its universality, considering systems with three, four, five, six, seven, eight, nine and ten species, described in three distinct square lattices, with $N=250, 500$ and $1000$.

The Hamming distance measures the difference between two final states, that evolve in time starting from two slightly different initial states, that differ from each other by a single site. However, due to the impossibility to directly control the randomness of the simulations, we elaborated the following strategy to calculate the Hamming distance: one generates an initial state and then makes a copy of it. One uses the initial state to run the simulation to get to the final state, which is saved. The key point here is that during the time evolution a new file is created, in which one saves every single step used to run it. One then takes the copy of the initial state and randomly selects a lattice site and modifies its content. When compared to the previous state, this new state has the tiniest difference, since among the $N^2$ sites in the square lattice it has a single site which is different. This new initial state is then used to run the same simulation already considered, evolving it according to the very same rules, in the same order and pace, as they appear in the saved file. The procedure leads to another final state, which is also saved.

The two final states are different, but the difference has nothing to do with the randomness of the stochastic simulations, being due to the tiniest modification introduced in the second initial state. To infer the presence of chaos, one then measures the difference between the two final states, generated with the same stochastic rules. Toward this goal, we employ the Hamming distance \cite{H} to measure the difference between the two final states. In fact, instead of the Hamming distance, one uses the Hamming distance density $h(t)=H(t)/N^2$, and displays the results using the average in a set of 1000 distinct values, that we collect from 1000 distinct simulations. The results are all shown in Fig.~\ref{fig3}, for the systems with three distinct lattice sizes, with $N=250, 500$ and $1000$. Since the two initial states have the tiniest difference due to the modification done in a single site in the lattice, we get $h(0)=1/N^2$, which practically vanishes for the lattice sizes considered in the current simulations.

% fig 4 %%%%%%%%%%%%%%%%%%%%%
\begin{figure}[t]
\centering
\includegraphics[scale=0.88]{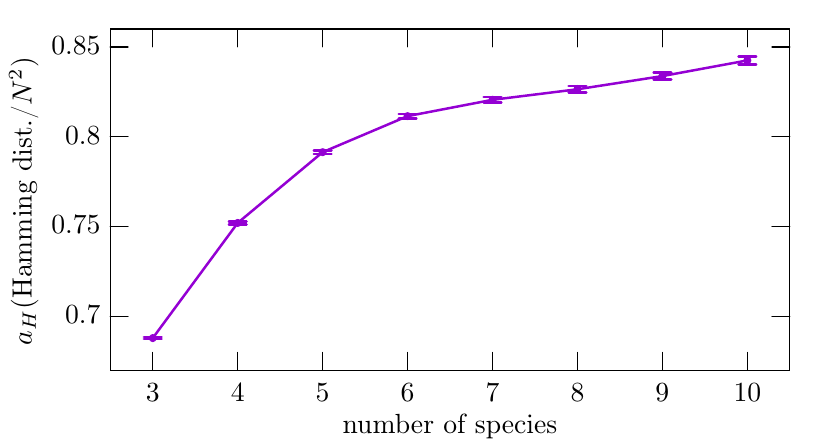}
\caption{The amplitude $a_H$ of the Hamming distance density as a function of the number of species for the lattice with $N=500$.}
\label{fig4}
\end{figure}
% fig 4 %%%%%%%%%%%%%%%%%%%%%

% fig 5 %%%%%%%%%%%%%%%%%%%%%
\begin{figure}[t]
\centering
\includegraphics[scale=0.88]{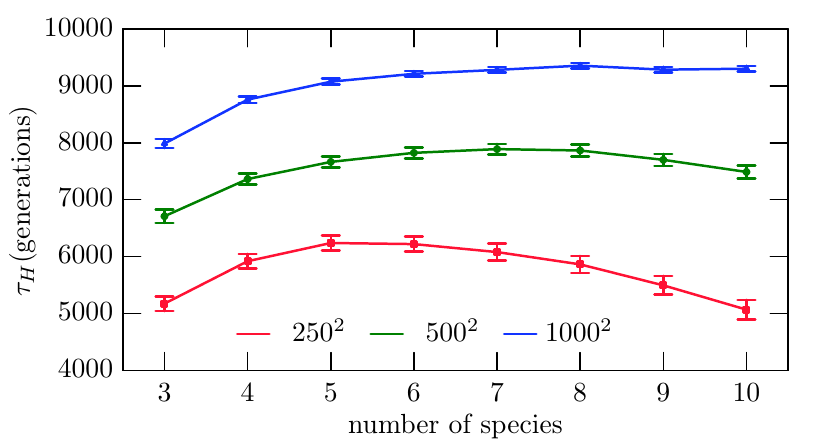}
\caption{The time $\tau_H$ necessary for the Hamming distance density to reach $a_H/2$, as a function of the number of species for the three distinct lattice sizes, with $N=250, 500$ and $1000$.}
\label{fig5}
\end{figure}
% fig 5 %%%%%%%%%%%%%%%%%%%%%

The results show that the Hamming distance density increases smoothly and then converges to a given value inside an interval with very narrow width in the square lattice. It has an universal behavior: it saturates at an average value that increases as one increases the number of species, although the system is always evolving in time as a non-equilibrium state with all the species fluctuating around an average value \cite{SR}. This is a consequence of the fact that as we increase the number of species, each species predates an increasing number of distinct species, meaning that the larger the number of species, the more likely the change of one single site to induce a larger alteration in the network configuration. In addition, we also calculate $\tau_H$, which is the time spent for the Hamming distance density to reach $a_H/2$ as one increases the number of species and the size of the lattice. We then see that the Hamming distance density $h(t)$ has amplitude $a_H$ and width $\tau_H$ that depend on the size of the lattice and the number of species. Since the amplitude $a_H$ is almost insensitive to the size of the lattice, we depict it in Fig.~\ref{fig4} for the lattice size $N=500$. The width is different, and it is displayed in Fig.~\ref{fig5} for three lattice sizes, as a function of the number of species.

% fig 6 %%%%%%%%%%%%%%%%%%%%%
\begin{figure}[t]
\centering
\includegraphics[scale=0.88]{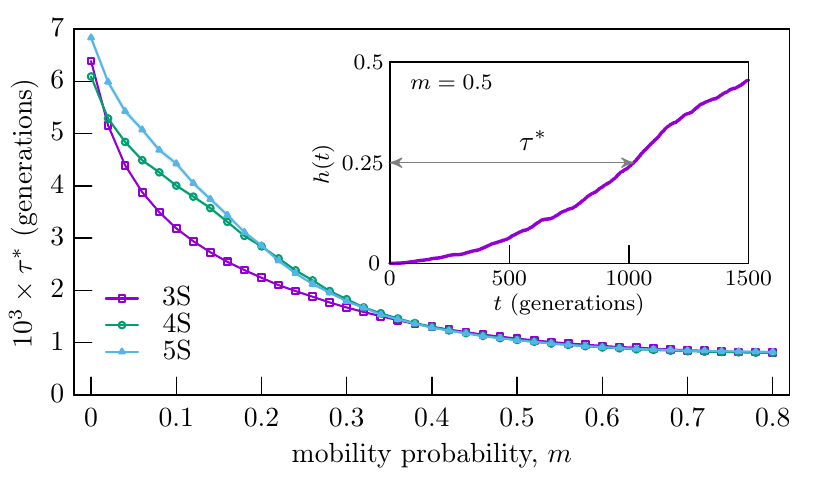}
\caption{Number of generations necessary for the Hamming distance density to reach the value $0.25$, as a function of the mobility. We considered a lattice with $N=500$ for the systems with three, four, and five species, and used an average of 1000 simulations with distinct initial conditions. In the inset one shows the result for a single realization, using the system with three distinct species and $m=0.5$.}
\label{fig6}
\end{figure}
% fig 6 %%%%%%%%%%%%%%%%%%%%%

In order to further explore the behavior of the Hamming distance density, we introduce $\tau^{*}$ as the time necessary for the two distinct evolutions to differ from each other by $1/4$, which is the same as the time spent for the Hamming distance density to reach the value $0.25$. We have noted that if one increases the mobility, the time $\tau^{*}$ diminishes, as expected, since the individuals can reach longer distances in shorter times. We have also noted that the behavior does not depend on the number of species for larger and larger values of $m$. The results are displayed in Fig.~\ref{fig6} for a lattice with $N=500$, in the case of systems with three, four, and five species. The behavior of $\tau^*$ remind us of the more conventional approach that deals with the Lyapunov exponent. This issue deserves further investigation, and we hope to report on it in another work. In particular, it would be of current interest to compare the Hamming distance density that we use in this work with other procedures, as the ones presented in Refs.~\cite{r1,r2}, for instance. In the work \cite{r1} the authors considers a continuum model of the Lotka-Volterra type, and this was also studied in \cite{baz}. One sees in Fig.~6 of Ref.~\cite{baz} that the stochastic simulations and the mean field simulations that come from the Lotka-Volterra approach are quite similar, so one thinks that stochastic simulations and the numerical Lotka-Volterra studies would lead to similar results.

\begin{figure}[t]
\centering
\includegraphics[scale=0.8]{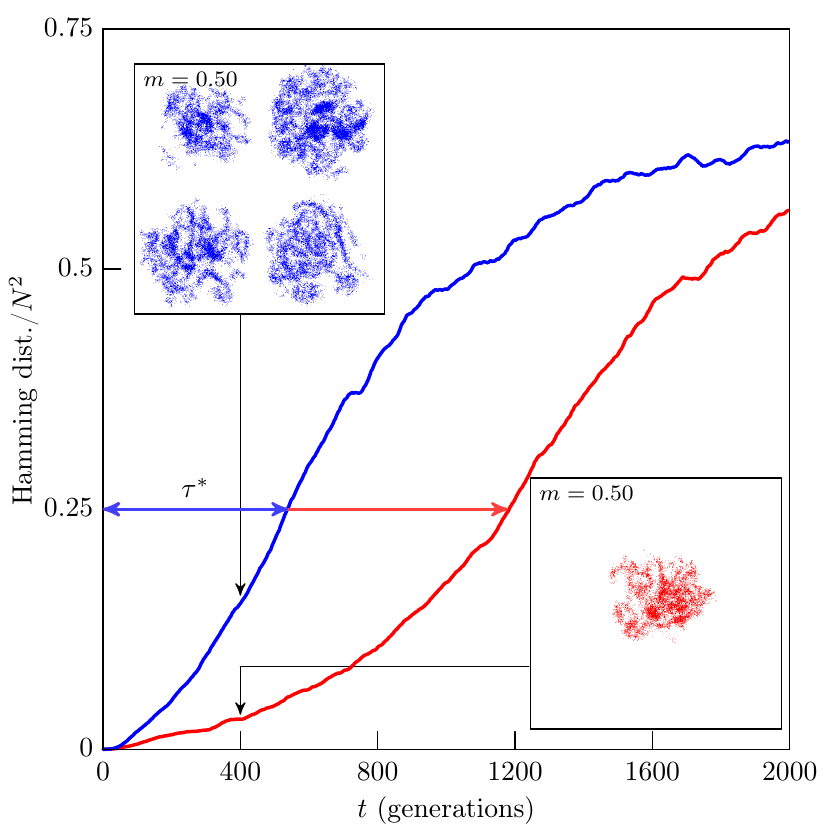}
\caption{The Hamming distance density for the blue and red evolutions, with the simulations being carried out with $m=0.50$, from distinct initial states, with four distinct lattice points being equally spaced (the blue case in the top inset) or grouped at the center of the lattice (the red case in the bottom inset).}
\label{fig7}
\end{figure}

In Fig.~\ref{fig7} one depicts stochastic simulations for three species and two distinct evolutions with $m=0.50$ and four distinct lattice points, one in blue, which shows the Hamming distance density with the two initial states differing by four distinct and equally spaced points, and the other, in red, with the four points grouped at the center of the lattice. We illustrate this in Fig.~\ref{fig7}, with the top and bottom insets displayed in blue and in red, respectively.  The results show that if the four points are left together, the total number of neighbor sites in the lattice decreases from 32 to 12, and this slows down the dispersal of information throughout the lattice, resulting in a larger $\tau^*$. We are then led to the conclusion that the higher the mobility, the faster the Hamming distance density reaches its maximum amplitude.

% fig 7 %%%%%%%%%%%%%%%%%%%%%
% fig 7 %%%%%%%%%%%%%%%%%%%%%

%%%%%%%%%%%%%%%%%%%%%%%%%%%%%%%%%%%%%%%%%%%%%%%%%%%%%%%%%%%%%%%%%%%%%%
\section{Mobility}
\label{sec4}

The above results suggest that the Hamming distance density has the universal behavior shown in Fig.~\ref{fig3} when the system engenders biodiversity. In this sense, it is of current interest to investigate the importance of mobility for the preservation of diversity, and how this can be related with the Hamming distance. The issue here is similar to the case developed in \cite{n07}, but one notes that it has been studied more recently in several works, in particular in \cite{M1,M2,M3,M4,M5} with distinct motivation. The purpose of this section is then to study the extinction of diversity with focus similar to the cases presented before in \cite{n07,M2,M4}. Here, however, one deals with three systems, with three, four and five distinct species, defined on two distinct lattices. Also, one considers the Moore neighborhood and predation in the form displayed in Fig.~\ref{fig1}, which differs from some previous investigations.

\begin{figure}[t]
\centering 
\includegraphics[scale=1.0]{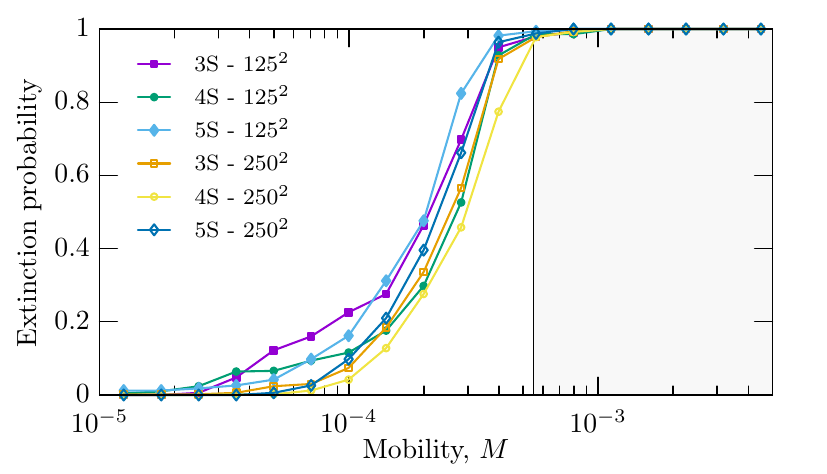}
\caption{Extinction probability as a function of $M$ for the three systems with three, four and five species, for the lattice with size $N=125$ and $N=250$.}
\label{fig8}
\end{figure}

To implement the investigation one first relaxes the constraint $m+p+r=1$ and use $p'=1=r'$, so they are not normalized anymore.  However, we can make them normalized using the following probabilities for mobility, reproduction, and predation or competition $m'/(m'+2)$, $1/(m'+2)$, and $1/(m'+2)$, respectively. According to the random walk theory, we can introduce another parameter, $M=m'/2N^2$, and make it the mobility, now being naturally proportional to the typical area explored by an individual per unit time. We then investigate the extinction probability, that is, the probability of extinction of diversity as a function of $M$. We display the results in Fig.~\ref{fig8} for the case of three, four and five species, for two lattices with $N=125$ and $N=250$, respectively. The vertical line shows the critical mobility, $M_c=(5.5\pm0.5)\,10^{-4}$, which is in good accordance with results obtained in Refs.~\cite{n07,M2,M4}.\\

% fig 8 %%%%%%%%%%%%%%%%%%%%%
% fig 8 %%%%%%%%%%%%%%%%%%%%%

%%%%%%%%%%%%%%%%%%%%%%%%%%%%%%%%%%%%%%%%%%%%%%%%%%%%%%%%%%%%%%%%%%%%%%
\section{Hamming distance versus mobility}
\label{sec5}

Since the Hamming distance density unveils the chaotic behavior, we can then investigate how it behaves as we vary the mobility. This is an alternative way to verify the presence of biodiversity as a function of mobility. To investigate this issue, we vary mobility keeping reproduction and predation or competition obeying $p'=r'=(1-m')/2$, and we describe the Hamming distance density as a function of the generation time for some specific values of the mobility. We display the results in Fig.~\ref{fig9}, and there one sees that for $m$ below the critical mobility, it keeps the universal behavior shown before in Fig.~\ref{fig3}. However, for $m$ above the critical value, it behaves differently, with a profile that shows the end of diversity as time goes by. To better understand this, one recalls the calculation of the Hamming distance: since we are comparing two states, when diversity is present, the Hamming distance has to relax to a given value inside the open set $(0,1)$, as it appears in Fig.~\ref{fig3}; however, when diversity is destroyed, the Hamming distance density has to become zero or unit, because the two final states may contain the same species, or two distinct ones, respectively. This behavior appears in the bottom panel in Fig.~\ref{fig9}, where we have simulated the three cases of $m=0.97$, $0.98$ and $0.99$, above the critical mobility. This is the first time the Hamming distance concept is used to confirm the end of biodiversity, as the mobility increases to larger and larger values.

The Hamming distance densities displayed in the top panel of Fig.~\ref{fig9} oscillate in a way which is not present in Fig.~\ref{fig3}. This happens because in Fig.~\ref{fig9} one shows a single simulation, and in Fig.~\ref{fig3} it is displayed the average over 1000 simulations, which smooths the corresponding curves.  Evidently, here in Fig.~\ref{fig9} a single simulation is enough to see if the Hamming distance density maintain its universal behavior which ensures biodiversity, or if it changes drastically, indicating the end of biodiversity.

%%%%%%%%%%%%%%%%%%%%%%%%%%%%%%%%%%%%%%%%%%%%%%%%%%%%%%%%%%%%%%%%%%%%%%
\section{Ending comments}
\label{sec6}

In this work we studied how the Hamming distance behaves for systems with three, four, five, six, seven, eight, nine and ten species, defined on square lattices of $250\times250$, $500\times500$ and $1000\times1000$ sites. The results show that the Hamming distance density has an universal profile, with amplitude $a_H(n)$ that depends mainly on the number of species $n$, and width $\tau_H(N, n)$ that depends on the lattice size $N$ and on the number of species, $n$.

\begin{figure}[t]
\centering
\includegraphics[scale=0.88]{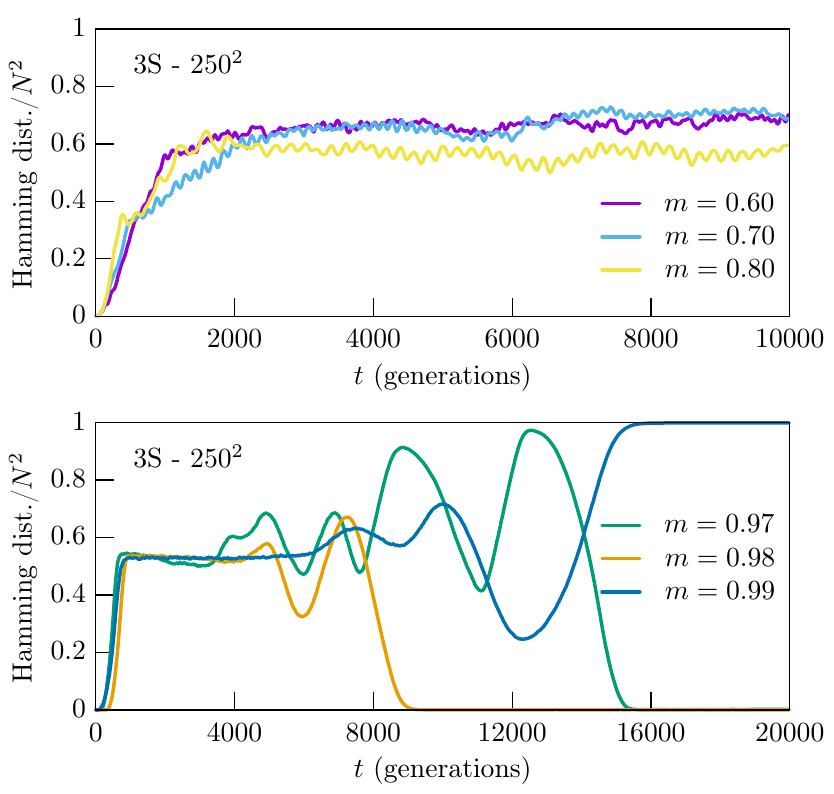}
\caption{The Hamming distance density as a function of the generation time, calculated for three distinct values of mobility. In the top panel the mobilities are below the critical mobility, and in the bottom panel they are above the critical mobility. We considered the system with three distinct species, and used $N=250$.}
\label{fig9}
\end{figure}

The time evolution shows that the Hamming distance density increases and reaches its maximum with amplitude $a_H(n)$ that is in between zero and unit, after a large number of generations. It has the universal behavior which is depicted in Fig.~\ref{fig3} for systems that maintain biodiversity, but if mobility is high enough to jeopardize biodiversity, the behavior changes drastically, as it is illustrated in Fig.~\ref{fig9}. The results displayed in Fig.~\ref{fig9} show that the Hamming distance density, which was recently used in \cite{SR} to unveil the chaotic behavior of stochastic simulations, can also be used as a way to measure the presence of biodiversity in models such as the ones studied in the current work.

% fig 9 %%%%%%%%%%%%%%%%%%%%%
% fig 9 %%%%%%%%%%%%%%%%%%%%%

The Hamming distance density concept that we studied in this work can be used to investigate rules that jeopardize biodiversity, to see if they change the Hamming distance behavior when they are added to the set of rules that control the stochastic evolution. Particularly interesting possibilities have been recently studied in Refs.~\cite{n1,n2,n3,n4,n5}, and the systems may perhaps develop behavior similar to the ones shown in Fig.~\ref{fig9}, if they can be controlled to cease biodiversity. We can also investigate the Hamming distance density starting with two identical copies of the initial state, but using different stochastic evolutions. Another issue concerns the behavior of the Hamming distance density in a three-dimensional cubic lattice, to see if the dimension of the lattice may affect the behavior found in Fig.~\ref{fig3} in the case of the two-dimensional square lattice. These and other related issues are currently under consideration, and we hope to report on them in the near future.

\acknowledgments
This work was partially financed by the CNPq Grants 455931/2014-3, 306614/2014-6, 308241/2013-4 and 479960/2013-5. BFO acknowledges support from INCT-FCx.


\begin{thebibliography}{0}
\bibitem{b1}M.A. Nowak, Evolutionary Dynamics: Exploring the Equations of Life. Harvard University Press, 2006.
\bibitem{b2}M.A. Pinsky and S. Karlin, {\it An \revision{I}ntroduction to Stochastic Modeling}. Academic Press, 2011.
\bibitem{ML}R.M. May and W.J. Leonard, SIAM J. Appl. Math. 29, 243 (1975).
\bibitem{rps}B. Sinervo and C.M. Lively, Nature 380, 240 (1996).
\bibitem{kd}T. Killingback and M. Doebeli, Proc. Roy. Soc. B {\bf263}, 1135 (1996).
\bibitem{DL}R. Durrett and S. Levin, Theor. Pop. Biol. 53, 30 (1998).
\bibitem{Sch}P. Schippers, A. M. Verschoor, M. Vos, and W. M. Mooij, Ecology Letters, {\bf4}, 404 (2001).
\bibitem{n02}B. Kerr, M.A. Riley, M.W. Feldman, and B.J.M. Bohannan, Nature 418, 171 (2002).
\bibitem{n04}B.C. Kirkup and M.A. Riley, Nature 428, 412 (2004).
\bibitem{n07}T. Reichenbach, M. Mobilia, and E. Frey, Nature 448, 1046 (2007).
\bibitem{S}G. Szab\'o and G. F\'ath, Phys. Rep. 446, 97 (2007).
\bibitem{SR}D. Bazeia, M.B.P.N. Pereira, A.V. Brito, B.F. de Oliveira, and J.G.G.S. Ramos, Sci. Rep. {\bf7}, 44900 (2017).
\bibitem{H}R.W. Hamming, Bell System Technical Journal, 29, 147 (1950). DOI:10.1002/j.1538-7305.1950.tb00463.x
\bibitem{gsl}
	\Name{Brian Gough},
	\Book{GNU Scientific Library Reference Manual},
	\Year{2009},
	\Publ{Network Theory Ltd.}
\bibitem{AB}P.P. Avelino, D. Bazeia, L. Losano, J. Menezes, and B.F. Oliveira, Phys. Rev. E {\bf86}, 036112 (2012).
\bibitem{r1}L. Roques, M.D. Chekroun, Ecological Complexity {\bf8}, 98 (2011).
\bibitem{r2}E. Beninc\`a, B. Ballantine, S. P. Ellner, and J. Huisman, PNAS {\bf212}, 6389 (2015).
\bibitem{baz}P. P. Avelino, D. Bazeia, J. Menezes, and B. F. de Oliveira, Phys. Lett. A {\bf378}, 393 (2014).
\bibitem{M1}S. Venkat and M. Pleimling, Phys. Rev. E {\bf81}, 021917 (2010).
\bibitem{M2}L.-L. Jiang, T. Zhou, M. Perc, and B.-H. Wang, Phys. Rev. E {\bf84}, 021912 (2011).
\bibitem{M3}A. Szolnoki et al., J. R. Soc. Interface {\bf11}, 20140735 (2014).
\bibitem{M4}H. Cheng et al., Sci. Rep. {\bf4}, 7486 (2014).
\bibitem{M5}M. Mobilia, A.M. Rucklidge, and B. Szczesny, Games {\bf7}, 24 (2016).
\bibitem{n1}J. Juul, K. Sneppen and J. Mathiesen, Phys. Rev. E {\bf87}, 042702 (2013).
\bibitem{n2}J. Vukov, A. Szolnoki, and G. Szab\'o, Phys. Rev. E {\bf88}, 022123 (2013).
\bibitem{n3}A. Szolnoki, J. Vukov, and M. Perc, Phys. Rev. E {\bf89}, 062125 (2014).
\bibitem{n4}A. Szolnoki and M. Perc, Phys. Rev. E {\bf93}, 062307 (2016).
\bibitem{n5}A. Szolnoki and M. Perc, Sci. Rep. {\bf6}, 38608 (2016).

\end{thebibliography}
\end{document}